# The Bunch Injection Controller for the PEP-II Storage Rings

R. Chestnut, A. Fisher, SLAC, Stanford, CA 94025, USA

Abstract

The PEP-II storage rings at SLAC each have 3492 'buckets' into which electrons and positrons can be injected into the high- and low-energy rings. Equipment to measure the currents of all the individual buckets was originally provided by the Lawrence Berkeley Laboratory and is implemented in VXI-based hardware. Data from this equipment as well as high precision direct current measurement provide the hard data for the Bunch Injection Controller. A large number of parameters determined by injection design considerations as well as set by operators for different circumstances are also used by the software algorithms to determine the desired bucket injection order and charge quantity for each injection pulse. These requests are then passed on to the venerable SLAC master pattern generator, which provides beams for other applications as well. This highly visible and highly successful system is implemented using the EPICS toolkit, and fits well into the merged SLAC EPICS/SLC control system. The Bunch Injection Controller hardware is a VME-based EPICS IOC, which makes extensive use of shared memory for communicating with the VXI measurement equipment and the SLAC master pattern generator.

## 1 PEP-II INJECTION REQUIREMENTS

The individual 3492 "buckets" in each ring circulate at a frequency of about 136 kHz, and are separated by 2.1 nanoseconds. The SLAC Linac can inject into both rings at 30Hz simultaneously or one ring at 60Hz. There are four "quanta" or injection chunks selectable for each individual injection request. The largest is about $10^{10}$ particles, corresponding to an increase in current of about 218 microamperes; smaller quanta are approximately 2/3 of the next larger.

The most important constraints on injection are:
- Allow arbitrary fill patterns.
- Observe the bunch damping times of 35ms (LER) and 60ms (HER) when considering where to inject next.
- Fill each bucket to within 2% of the requested amount where possible.
- Fill the rings as evenly as possible.
- NEVER overfill a bucket.

## 2 PRE-EXISTING HARDWARE

### 2.1 Bunch-by-Bunch Current Monitor (BxBCM)

LBL was in the process of implementing hardware when the injection requirements were under discussion. The BxBCM implementation consists of analog down-converters (by Jim Hinckson) and a set of VXI-based multiplexers and 8-bit fast ADCs, controlled by a PC-based slot-0 controller running NT (by Mike Chin). Each ring has its own analog and VXI support. This equipment digitizes signals for each bucket 250 times each 1/60 second. The NT-based software prepares one second sums for each bucket. There is, additionally, control and read-out of attenuations and crate health.

### 2.2 Direct Current Current Transformer (DCCT)

For each ring the pre-existing DCCT provides a stable voltage derived from the total ring current. These voltages are each digitized twice a second by individual Keithley 2002 Digital Volt Meters.

### 2.3 Master Pattern Generator (MPG)

The MPG is a well-established iRMX-based machine, which has been running the SLAC beams for two decades. Since a secondary requirement foresees interleaved operation between PEP-II injection and other linac operations, the injection implementation needed to fit seamlessly into the old system.

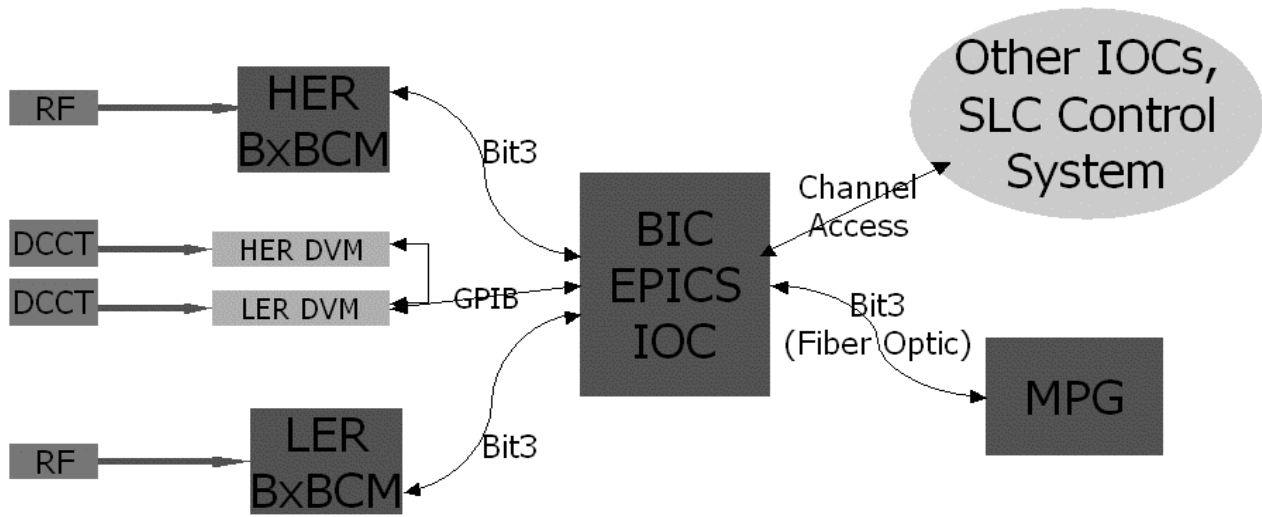

**Figure 1: Cartoon of Bunch Injection Controller Interfaces**

## 3 IMPLEMENTATION

We chose to implement the Bunch Injection Controller (BIC) as an EPICS VME-based IOC using shared memory to communicate with the pre-existing computers and GPIB for the Kiethley DVM readout. The first SLAC EPICS project, PEP-II RF control, had been very successful and we were enthusiastic about further use of EPICS. The BIT-3 shared memory provides access with the lowest degree of coupling and fewest software and hardware changes to the pre-existing subsystems. The dual-port RAM itself is resident in the two BxBCM VXI crates and in the SLAC MPG so that these systems continue to run independently of the BIC, the new system under development.

### 3.1 Bunch Current Monitor BIT-3 Interface

Each BIT-3 memory is partitioned into a small control segment, a small diagnostic read-back segment, and a large data array segment. Pre- and post-data array update counters are used to guarantee that the BIC reads consistent data. These counters also serve as a heartbeat. There is no other synchronization.

### 3.2 Master Pattern Generator BIT-3 Interface

The BIT-3 memory is dedicated to two small ring buffers. One contains request information written by the BIC for the MPG and the other contains responses and status from the MPG for the BIC. Both the BIC and MPG code require no synchronization.

**BIC => MPG requests**

| Index | Quantum | Bucket | Delta Q |
|---|---|---|---|

The **Delta Q** is actually filled in later, by comparing the stable bucket current before injection with stable current after injection. This incremental charge is used to maintain a dynamic calibration of the injection quanta.

**MPG => BIC responses**

| Index | Pulse-ID | Status |
|---|---|---|

The **Pulse-ID** in the MPG response is used by monitor code in conjunction with **Pulse-ID**-tagged Toroid measurements in the injection lines for efficiency calculations.

### 3.3 Kiethley DVM GPIB interface

One GPIB controller serves both the LER and HER Digital Volt Meters. The DCCT signals are preprocessed so that 5 Volts corresponds to 10 Ampere in the ring. Each DVM is read out at 2 Hz. The dI/dT calculations are done in software in lieu of reading those data from the DVMs as well. This read-out is not synchronized to the bucket-by-bucket information. The time stamp from the DVM is used only for diagnostics.

## 4 BUNCH CURRENT PROCESSING

### 4.1 Hardware Processing

The signal from the beam pick-off is mixed with an RF carrier to form two periods of an amplitude modulated (approximate) sine wave. Two "decimator" modules alternate between the even and odd buckets each turn, feeding the data into fast 8-bit ADCs. Each of the 3492 individual buckets is digitized 250 times each 1/60 second. The data are then copied from the VXI module into the Slot-0 Controller computer memory. After every data accumulation, the phase of the incoming signal is shifted 90 degrees.

## 4.2 Software Processing

The computer code accumulates two 3492 bucket arrays, one containing 0 degree data minus 180 degree data, the other containing (90-270) degree data. This subtraction removes the background and provides clean "sine" and "cosine" phase data. Each second, after taking 15 samples at each phase, the data are transferred to the shared memory, where the BIC can access the two arrays.

In the BIC, these data are used to generate the amplitudes and phases of each bucket. Since the hardware phase shifter does not shift a true 90 degrees, to first order the actual amplitude is given by:

**Amplitude=sqrt(S*S+C*C-2*S*C*delta),**

Where delta is the average deviation of the phase shifter from 90 degrees, S is the "sine" value, and C the "cosine" value. This correction provides a wide flat region where the amplitude is insensitive to the phase setting. The value of **delta** is calculated by measuring **S**, **C**, and **amplitude** over a range of phase settings and finding a **delta** that provides the widest flat response.

These data are then normalized to the total ring current and the resulting array of microampere per bucket is used in subsequent processing.

## 4.3 Filling Algorithm

The fill goal is specified by a pattern description and the total desired ring current. Simple pattern descriptions can be typed directly into a control variable, or complex patterns can be entered into a data file, pointed to by the control variable. The pattern "language" provides shortcuts for specifying whole ranges of buckets and for easily specifying a "ramp", where the desired fill value increases smoothly over a range of buckets. Combining the pattern and total desired ring current creates an array of desired bucket currents (microampere).

There are four different injection chunks (quanta) available for injection into each ring, each about 2/3 of the next larger. These allow for fast bulk injection and "fine tuning" near the end.

To provide a smooth fill, intermediate goals of 1.2, 2.2, etc. times the value of the largest quanta are generated in turn as the previous intermediate goal is filled. Only the largest quanta are used in this phase. At the end of filling, when using the true goal, all quanta are considered.

A candidate bucket is checked to verify that an injection will not overfill. If a candidate qualifies, a request is placed on the queue to the Master Pattern Generator (MPG). The next bucket considered is at least 197 buckets distant, so that the newly injected bucket has time to damp to the nominal beam size. If a candidate does not qualify, the next adjacent bucket is considered. Each second only sufficient requests are made to the MPG queue so that the MPG is never idle. Right near the end of filling, spacers are often inserted in the queue to avoid filling buckets too closely spaced.

This approach is very successful in filling empty rings, topping off, and in filling with many "drop outs", i.e. buckets which lose disproportionately high current due to tune or beam-beam interaction problems.

## 4.4 Efficiencies and Calibration

The four individual injection quanta are dynamically calibrated by observing the change in current for each satisfied injection request. Each measurement is passed through an anomalous value filter and used as input for a simple smoothing function. Further, the injection efficiency is calculated by comparing injection line toroid measurements to measured bucket current increases, using the MPG Linac pulse-id as the correlator, as described above. The numerology describing the Linac damping rings and PEP-II ring result in a difference of four buckets for each damping ring turn. For a difference of fewer than four buckets, the Linac is rephased. Therefore injection efficiency is separately calculated for each bucket number modulo four.

## 4.5 Control Interface and Displays

Injection is controlled through a standard EPICS Display Manager control interface. Detailed and expert panels are available as a tree descending from the main panel. There are some standard control room overhead displays of daily luminosity and currents as well as expanded currents and lifetimes for the last twenty and two minutes. The most useful for filling, however, is a display of actual bucket currents normalized to the goal currents. After a successful fill, all visible points cluster near a value of unity. Buckets not represented in the goal pattern are not visible, unless they actually have some current, in which case those buckets have the value 1.2, noticeably above the normal fill. Thus timing problems, resulting in filling the wrong bucket, are easily visible.

## 5 EXPERIENCES AND PROBLEMS

This system was ready at first beam in PEP-II and commissioned soon thereafter. Three distinct types of problems were noted:
1) GPIB malfunction – multiplexing one digital voltmeter for both rings caused many deadlocks. We now use two digital voltmeters.
2) Bad cables – loose or intermittent BIT-3 cables caused difficult problems.
3) Timing shifts – by far the most frequent problem. Evolving diagnostic displays now enable quick operational response.

As we have solved problems, the symptoms, diagnostic method, and solutions have been added to help pages on the web. Operations personnel now routinely diagnose complex injection problems on their own.

## REFERENCES

[1] V. Bharadwaj et al., "PEP-II Injection Timing and Controls", SLAC-PUB-7600, Jul 1997. Talk given at 17$^{th}$ IEEE Particle Accelerator Conference (PAC 97), Vancouver, Canada, 12-16 May 1997.

[2] Gregory A. Lowe, "Flexibility of the PEP-II Injection Scheme", SLAC-PEP-II-AP-NOTE-3-94, Feb 1994.

[3] Fieguth, et al., "Injection System for the PEP-II Asymmetric B Factory at SLAC", SLAC-PUB-5775, Mar 1992. Also Berlin 1992, Proceedings, EPAC92, vol. 2 pp. 1443-1445, presented at the 3$^{rd}$ European Particle Accelerator Conference, Berlin, Germany, Mar 24-28, 1992.